\documentclass[12pt]{article}
\usepackage{amsmath}
\usepackage{amssymb}
\usepackage{amsthm}
\usepackage{bm}
\usepackage{graphicx}
\usepackage{subfig}
\usepackage{cite}
\usepackage[usenames,dvipsnames]{xcolor}
\usepackage[hidelinks]{hyperref}
\usepackage[capitalise]{cleveref}

\newcommand{\Vector}[1]{\mathbf{#1}}
\newcommand{\Matrix}[1]{\mathbf{#1}}

\title{The Replicator Dynamics of Zero-Sum Games Arise from a Novel Poisson Algebra}

\author{Christopher Griffin\footnote{Applied Research Laboratory, Penn State University, University Park, PA 16802, E-mail: griffinch@psu.edu}}

\begin{document}
\maketitle

\begin{abstract} We show that the replicator dynamics for zero-sum games arises as a result of a non-canonical bracket that is a hybrid between a Poisson Bracket and a Nambu Bracket. The resulting non-canonical bracket is parameterized both the by the skew-symmetric payoff matrix and a mediating function. The mediating function is only sometimes a conserved quantity, but plays a critical role in the determination of the dynamics. As a by-product, we show that for the replicator dynamics this function arises in the definition of a natural metric on which phase flow-volume is preserved. Additionally, we show that the non-canonical bracket satisfies all the same identities as the Poisson bracket except for the Jacobi identity (JI), which is satisfied for special cases of the mediating function. In particular, the mediating function that gives rise to the replicator dynamics yields a bracket that satisfies JI. This neatly explains why the mediating function allows us to derive a metric on which phase flow is conserved and suggests a natural geometry for zero-sum games that extends the Symplectic geometry of the Poisson bracket and potentially an alternate approach to quantizing evolutionary games. 
\end{abstract}

\section{Introduction}
Evolutionary games have been well-studied over the past four decades \cite{AL84,W97,HS98,HS03,T15,T19}, with much attention paid to the replicator equation. The Hamiltonian structure of these systems has been considered in \cite{Hof96} and also in \cite{SAF02,SC03,SAC05}, where chaotic motion in generalized replicators are also considered. There has also been recent interest in the intersection of Lie Algebras, Lie Brackets and the replicator dynamic \cite{RK20,GF17}.

Nambu mechanics \cite{N73} are a logical generalization of Hamiltonian dynamics to odd dimensional systems and systems with multiple conserved quantities that are not (necessarily) defined by Hamilton's equations yet (i) are conservative and (ii) satisfy their own more generic form of Liouville's theorem. 
Nambu first defined his bracket on two Hamiltonians. Let $F,G,H:\mathbb{R}^3 \to \mathbb{R}$ be three functions. The Nambu bracket in which $G$ and $H$ are conserved quantities is
\begin{displaymath}
[F,G,H] = \left\langle{\nabla F , \nabla G \times \nabla H}\right\rangle,
\end{displaymath}
where $\langle{\cdot,\cdot}\rangle$ is the standard inner product.

A three dimensional dynamical system with variables $(x_1,x_2,x_3)$  has Nambu dynamics if
\begin{displaymath}
\dot{x}_i = [x_i, G, H]
\end{displaymath} 
for some conserved quantities $G$ and $H$. In general on $H_1,\dots,H_{n-1}$ conserved quantities the Nambu bracket becomes
\begin{equation}
\left[F,H_1,\dots,H_{n-1}\right] = \sum_{\sigma \in S_{n}} \epsilon_\sigma \frac{\partial F}{\partial x_{\sigma(1)}} \frac{\partial H_1}{\partial x_{\sigma(2)}}\cdots \frac{\partial H_{n-1}}{\partial x_{\sigma(n)}}.
\end{equation}
Here $\sigma$ is a permutation in the symmetric group $S_n$ on $n$ symbols and $\epsilon_{\sigma}$ is the Levi-Civita symbol at index $\sigma(1),\dots,\sigma(n)$. Nambu \cite{N73} observed that when $F = x_i$, the resulting system given by
\begin{equation}
\dot{x}_i = [x_i, H_1,\dots,H_{n-1}]
\label{eqn:Nambu}
\end{equation}
obeys a generalized form of Liouville's Theorem. Necessarily any $n$-dimensional system of ordinary differential equations with exactly $n-1$ conserved quantities must be of the form \cref{eqn:Nambu}. In what follows we consider dynamical systems that evolve on the $n-1$ dimensional unit simplex embedded in $\mathbb{R}^n$, which we denote $\Delta_{n-1}$. 

\section{Game Theoretic Motivation}
To motivate the discussion, consider the rock-paper-scissors (RPS) payoff matrix
\begin{equation}
\Matrix{A} = \begin{bmatrix}0 & -1 & 1\\1 & 0 & -1\\-1 & 1 & 0\end{bmatrix}.
\label{eqn:AMatrix0}
\end{equation}
Let $\Vector{x} = (x_1,x_2,x_3) \in \Delta_2$ be the proportions of rock, paper and scissors (resp.) in an evolutionary game. We assume $\mathbf{x}$ is a column vector with corresponding row vector $\mathbf{x}^T$. It is known that for RPS, the replicator dynamic \cite{EA83}
\begin{displaymath}
\dot{x}_i = x_i \left(\Vector{e}_i^T\Matrix{A}\Vector{x} - \Vector{x}^T\Matrix{A}\Vector{x}\right)
\end{displaymath}
conserves precisely the two quantities $G(\Vector{x}) = x_1 x_2 x_3$ and $H(\Vector{x}) = x_1 + x_2 + x_3$. Therefore, ordinary RPS exhibits a Nambu dynamic with
\begin{displaymath}
\dot{x}_i = [x_i,G,H].
\end{displaymath}
The conservation of the quantity $G(\Vector{x})$ reflects the conservation of volume in the Euclidean norm in the flow in phase space, though it is certainly not necessary that $G(\Vector{x})$ be conserved to have Euclidean volume conservation in phase space. \textcolor{black}{For example when
\begin{equation}
G(x) = -\sum_i x_i\log(x_i),
\label{eqn:GEntropy}
\end{equation} 
the dynamics conserve the Euclidean volume of phase space.} In this case, however, entropy is conserved and the dynamics become
\begin{equation}
\dot{x}_i = \mathbf{e}_i^T\mathbf{A}\mathbf{x}_{\log},
\label{eqn:SimpleLog}
\end{equation}
where $\mathbf{x}_{\log} = \langle{\log(x_1),\log(x_2),\log(x_3)}\rangle$.

\subsection{Results in Three Strategy Zero-Sum Games}
Consider the generalized zero-sum (skew-symmetric) payoff matrix
\begin{displaymath}
\Matrix{A} = \begin{bmatrix}
0 & -b & a\\
b & 0 & -c\\
-a & c & 0
\end{bmatrix}.
\end{displaymath}
This matrix has $0$ determinant so by Zeeman's theorem \cite{Ze80}, the interior of $\Delta_2$ has no hyperbolic fixed points. The phase flow does not preserve Euclidean volume. However, it is noted in \cite{EA83,Hof96} that for replicator dynamics with an elliptic interior fixed point, there is a 2-form dependent on the game matrix whose resulting volume is preserved; however the explicit structure of this two-form is never provided. In the case of the ordinary RPS matrix, this 2-form is the Euclidean metric tensor. Inspired by this, we can modify the Nambu bracket to accommodate a more generalized form of Liouville's theorem. Nambu's work was designed to generalize Liouville's theorem to cases when the dynamics are phase fluid volume preserving.

\begin{figure}[htbp]
\centering
\includegraphics[width=0.65\textwidth]{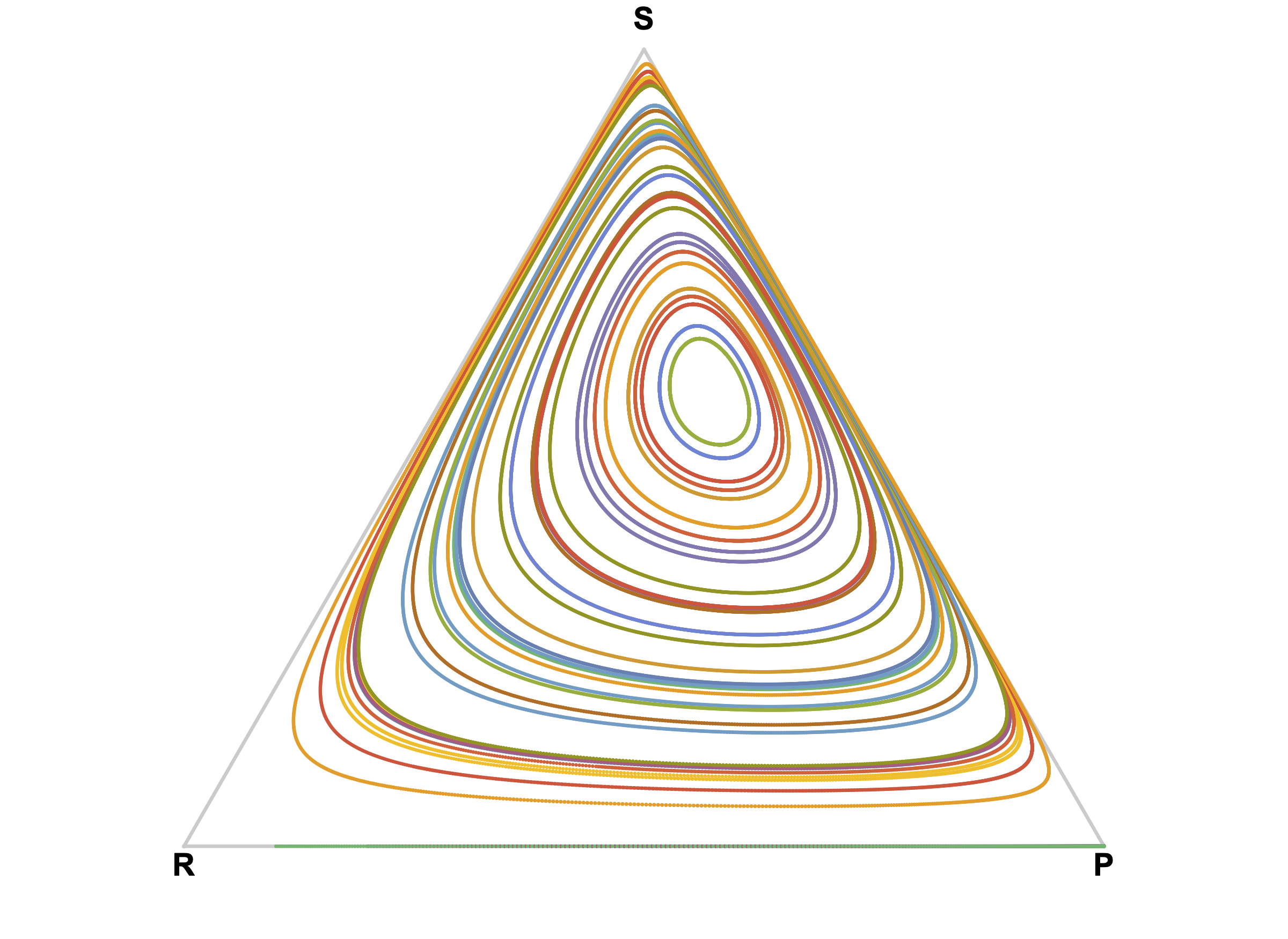}
\caption{The dynamics of \cref{eqn:3DQBracket} are shown. The visual asymmetry in the phase portrait is caused by the  $\Matrix{Q}$-modified Nambu bracket when $\mathbf{Q}$ is not a multiple of the identity. In this case, the standard Euclidean volume is not conserved by the dynamics. Here $a = 1$, $b = 2$ and $c = 1/2$.}
\label{fig:Asymmetry}
\end{figure}

Given a diagonal matrix $\Matrix{Q} \in \mathbb{R}^{3 \times 3}$, let the $\Matrix{Q}$-modified Nambu bracket be
\begin{displaymath}
[F,G,H]_Q = \left\langle{\nabla F , (\mathbf{Q}\cdot\nabla G) \times \nabla H}\right\rangle.
\end{displaymath}
If
\begin{displaymath}
\Matrix{Q} = \begin{bmatrix} c & 0 & 0 \\ 0 & a & 0 \\ 0 & 0 & b\end{bmatrix},
\end{displaymath}
then the replicator dynamics for the three strategy zero-sum RPS matrix are given by
\begin{equation}
\dot{x}_i = [x_i,G,H]_\mathbf{Q},
\label{eqn:3DQBracket}
\end{equation}
where $G$ and $H$ are defined as before. \textcolor{black}{When $\mathbf{Q}$ is not a multiple of the identity matrix, the interior fixed point of the system of differential equations moves from the center of $\Delta_2$ and the phase portrait becomes (visually) asymmetric (see \cref{fig:Asymmetry}).  As a further consequence, the Euclidean volume is no longer conserved.}

%There is a relationship between $\Matrix{Q}$ as defined above and a volume form that is conserved by the (generalized) rock paper scissors dynamics. We note this volume form is only valid on $\mathrm{int}(\Delta_3)$, which is to be expected since flow is never conserved on the boundary.

Let $g_R(\Vector{x}) = R/G(\Vector{x})$ (for some $R \in \mathbb{R}$). Let $\Vector{F}(\Vector{x})$ be the vector field defining the right-hand-side of the replicator dynamics for the (generalized) RPS game. Straightforward computation shows that in Euclidean coordinates
\begin{equation}
\nabla \cdot \left[g_R(\Vector{x}) \Vector{F}(\Vector{x})\right] = 0.
\label{eqn:ModifiedDivergence}
\end{equation}
Consequently, any metric tensor $\Matrix{g}$ with $\sqrt{\mathrm{det}(\Matrix{g})} = g_R(\Vector{x})$ will have the property that the divergence of the vector field $\Vector{F}(\Vector{x})$ with respect to this two-form is $0$. 
That is,
\begin{displaymath}
\textcolor{black}{\mathrm{div}\left[\Vector{F}(\Vector{x})\right]} = \frac{1}{\sqrt{\mathrm{det}(\Matrix{g})}} \sum_{i} \partial_{x_i}\left[\sqrt{\mathrm{det}(\Matrix{g})} \Vector{F}(\Vector{x})\right] = 0.
\end{displaymath}
This implies there is a family of metrics for which the divergence of the replicator dynamics vanishes. In particular, $\mathrm{div}\left[\Vector{F}(\Vector{x})\right]=0$ with respect to the  straightforward metric
\begin{equation}
\mathbf{g} = \begin{bmatrix}\tfrac{1}{x_1^2} & 0 & 0\\0 & \tfrac{1}{x_2^2} & 0\\0 & 0 & \tfrac{1}{x_3^2}
\end{bmatrix},
\label{eqn:Metric}
\end{equation}
which is similar to but distinct from the Shahshahani metric (see \cite{HS98}, \cite{RK20}). Before generalizing, it is worth noting that when $G(x)$ is replaced by \cref{eqn:GEntropy}, in \cref{eqn:3DQBracket}, the phase flow continues to preserve Euclidean volume. This seems to be an interesting property of the entropy in $\Delta_2$.

\section{Results in Arbitrary Dimensions}
Now consider an arbitrary dimension skew-symmetric $n\times n$ payoff matrix $\mathbf{A}$ with $n \geq 3$ defining a zero-sum game. We define a new bracket as
\begin{equation}
\{F,H\}_{\Matrix{A},G} = 
\sum_{\sigma \in S_n} \frac{\Matrix{A}_{\sigma(1),\sigma(n)}}{(n-2)!} \frac{\partial F}{\partial x_{\sigma(1)}} \frac{\partial^{n-2}G}{\partial x_{\sigma(2)}\cdots \partial x_{\sigma(n-1)}}\frac{\partial H}{\partial x_{\sigma(n)}}.
\label{eqn:Bracket}
\end{equation}
In the case when $n = 3$, this reduces to the Nambu bracket if and only if $\mathbf{A}$ is the three-strategy unbiased RPS game. One could argue with suitable abuse of notation that for $n = 2$, this yields a scaled version of the Poisson bracket, but we will not consider this case. Our demotion of $G$ out of the bracket will be clear in the sequel.

Because we assume $\mathbf{A}$ is skew-symmetric, this bracket has the anti-commutativity property
\begin{displaymath}
\{F,H\}_{\mathbf{A},G} = -\{H,F\}_{\mathbf{A},G}.
\end{displaymath}
From its construction, the bracket is bi-linear in the sense that for $\alpha,\beta\in\mathbb{R}$ we have
\begin{displaymath}
\{\alpha F_1 + F_2,H\}_{\mathbf{A},G} = \alpha\{F_1,H\}_{\mathbf{A},G} + \{F_2,H\}_{\mathbf{A},G}
\end{displaymath}
and
\begin{displaymath}
\{F,\beta H_1 + H_2\}_{\mathbf{A},G} = \beta\{F_1,H_1\}_{\mathbf{A},G} + \{F,H_2\}_{\mathbf{A},G}.
\end{displaymath}
Likewise, by the product rule, this bracket must satisfy a Leibniz rule
\begin{displaymath}
\{F_1F_2,H\}_{\mathbf{A},G} = \{F_1,H\}_{\mathbf{A},G}F_2 + F_1\{F_2,H\}_{\mathbf{A},G}.
\end{displaymath}
We delay discussing the Jacobi identify until after we have illustrated the utility of this bracket in evolutionary games.
 
For compactness, let
\begin{displaymath}
\frac{\partial^{n-2}G}{\partial x_{\sigma(2)}\cdots \partial x_{\sigma(n-1)}} = G_{{\sigma(2)}\cdots{\sigma(n-1)}}.
\end{displaymath}
%Let $H$ and $G$ be arbitrary and assume dynamics are given by \cref{eqn:BracketReplicator}. 
Then
\begin{displaymath}
\dot{H} = \sum_{i=1}^n \frac{\partial H}{\partial x_i}\dot{x}_i = 
\sum_{i=1}^n \sum_{\sigma \in S_n} \frac{\Matrix{A}_{\sigma(1),\sigma(n)}}{(n-2)!} \frac{\partial H}{\partial x_{\sigma(1)}} G_{{\sigma(2)}\cdots{\sigma(n-1)}}\frac{\partial H}{\partial x_{\sigma(n)}}.
\end{displaymath}
\textcolor{black}{With no loss of generality,} assume $\sigma(1) = i$ and $\sigma(n) = j$ \textcolor{black}{(arbitrary indices).} Define
\begin{equation}
T_{ij} = \frac{\Matrix{A}_{ij}}{(n-2)!}\frac{\partial H}{\partial x_{i}}\frac{\partial H}{\partial x_{j}}\sum_{\sigma \in S_n | \sigma(1) = i,\sigma(n) = j} G_{{\sigma(2)}\cdots{\sigma(n-1)}}.
\label{eqn:Tij}
\end{equation}
\textcolor{black}{By fixing $\sigma(1)$ and $\sigma(n)$, we can decompose $\dot{H}$ into a sum of the terms $T_{ij}$, which are easier to analyze. Effectively we are factoring out the various $G_{{\sigma(2)}\cdots{\sigma(n-1)}}$ for each possible value of $\sigma(1)$ and $\sigma(n)$.} Thus we see
\begin{displaymath}
\dot{H} = \sum_{i}\sum_{j} T_{ij}. 
\end{displaymath}
However, $T_{ij} = -T_{ji}$, \textcolor{black}{which follows from \cref{eqn:Tij} and the fact that $\mathbf{A}$ is skew-symmetric}. From this we conclude that $\dot{H} = 0$. Therefore, $H$ is a constant of motion in this modified bracket, as expected. Depending on $\mathbf{A}$, $G$ and $H$ it is possible that $G$ is conserved, but this is not guaranteed. We discuss this in the context of the standard replicator dynamics below.

Let
\begin{equation}
\quad G = \prod_i x_i, \quad H = \sum_i x_i.
\label{eqn:GReplicator}
\end{equation} 
Then the classic replicator dynamic on zero-sum games is given by
\begin{equation}
\dot{x}_i = \{x_i,H\}_{\mathbf{A},G}.
\label{eqn:BracketReplicator}
\end{equation}
To see this note that $\mathbf{x}^T\mathbf{A}\mathbf{x} = 0$. Therefore
\begin{equation}
\dot{x}_i = x_i \mathbf{e}_i^T\mathbf{A}\mathbf{x} = \sum_{j \neq i} \Matrix{A}_{ij}x_ix_j.
\label{eqn:ZeroSum}
\end{equation}
When $\sigma(1) = i$ and $\sigma(n) = j$, the coefficient of the summand is $\mathbf{A}_{ij}/(n-2)!$ and
\begin{equation}
G_{{\sigma(2)}\cdots{\sigma(n-1)}}= x_i x_j.
\label{eqn:xixj}
\end{equation}
This term is repeated $(n-2)!$ times because we sum over elements of the symmetric group. Thus expanding the right-hand-side of \cref{eqn:Bracket} yields \cref{eqn:ZeroSum} for the given $F$, $G$ and $H$. 

We also deduce that \cref{eqn:ModifiedDivergence} generalizes from the fact that $\mathbf{x}^T\mathbf{A}\mathbf{x} = 0$. From \cref{eqn:ZeroSum}, let $F_i(\mathbf{x}) = x_i \mathbf{e}_i^T\mathbf{A}\mathbf{x}$. Then
\begin{equation}
\left[\frac{1}{\prod_{j} x_j} F_i(\mathbf{x})\right] = \frac{\mathbf{e}_i^T\mathbf{A}\mathbf{x}}{\prod_{j \neq i} x_j},
\end{equation}
which has no terms in $x_i$ because $A_{ii} = 0$ by assumption. Therefore
\begin{equation}
\nabla \cdot \left[\frac{1}{G(\mathbf{x})}  \Vector{F}(\Vector{x})\right] = 0.
\label{eqn:ModifiedDivergence2}
\end{equation}
It follows that the volume of phase flow is preserved for the $n$-dimensional generalization of $\mathbf{g}$ in \cref{eqn:Metric}. This argument also allows us to produce a condition for Euclidean volume to be preserved. If the row (column) sum of $\mathbf{A}$ is zero, then Euclidean volume must be conserved since it is clear that
\begin{displaymath}
\frac{\partial}{\partial x_i} F_i(\mathbf{x}) = \mathbf{e}_i^T\mathbf{A}\mathbf{x},
\end{displaymath}
again because $\mathbf{A}_{ii} = 0$. Consequently, $\nabla\cdot\mathbf{F}(\mathbf{x}) = \mathbf{1}^T\mathbf{A}\mathbf{x}$, which is zero if and only if the row sum of $\mathbf{A}$ is zero. Here $\mathbf{1}$ is an n-dimensional vector of 1's. Modification to \cref{eqn:Tij} shows that a zero column (row) sum is also sufficient to guarantee that $G(\mathbf{x}) = x_1\cdots x_n$ is conserved. 

The preceding analysis suggests that the function $G$ should not be considered a conserved quantity (hence its demotion out of the bracket), but instead is a mediator of the functional form of the resulting dynamics. As shown, in the case when $n = 3$ and $G$ is the entropy of the discrete distribution defined by $\mathbf{x}$, the resulting dynamics are simple and governed by \cref{eqn:SimpleLog}. In the case when $n = 4$, if $G$ is defined as the entropy of $\mathbf{x}$, the resulting dynamics are trivial because $G_{{\sigma(2)}\cdots{\sigma(n-1)}} = 0$. However, if we define
\begin{equation}
G(x_1,x_2,x_3,x_4) = \sum_{i} x_i\prod_{j\neq i} \log(x_i),
\label{eqn:ScrewedUpEntropy}
\end{equation}
then just as in \cref{eqn:SimpleLog}, we have
\begin{displaymath}
\dot{x}_i = \{x_i,H\}_{\mathbf{A},G} = \mathbf{e}_i^T\mathbf{A}\mathbf{x}_{\log},
\end{displaymath}
and again this system of ordinary differential equations conserves Euclidean volume. For $n > 4$, a similar $G(\mathbf{x})$ function can be defined yielding \cref{eqn:SimpleLog}, but the interpretation becomes more difficult. 

\subsection{Jacobi Identity}
\begin{figure}[htbp]
\centering
\includegraphics[width=0.45\textwidth]{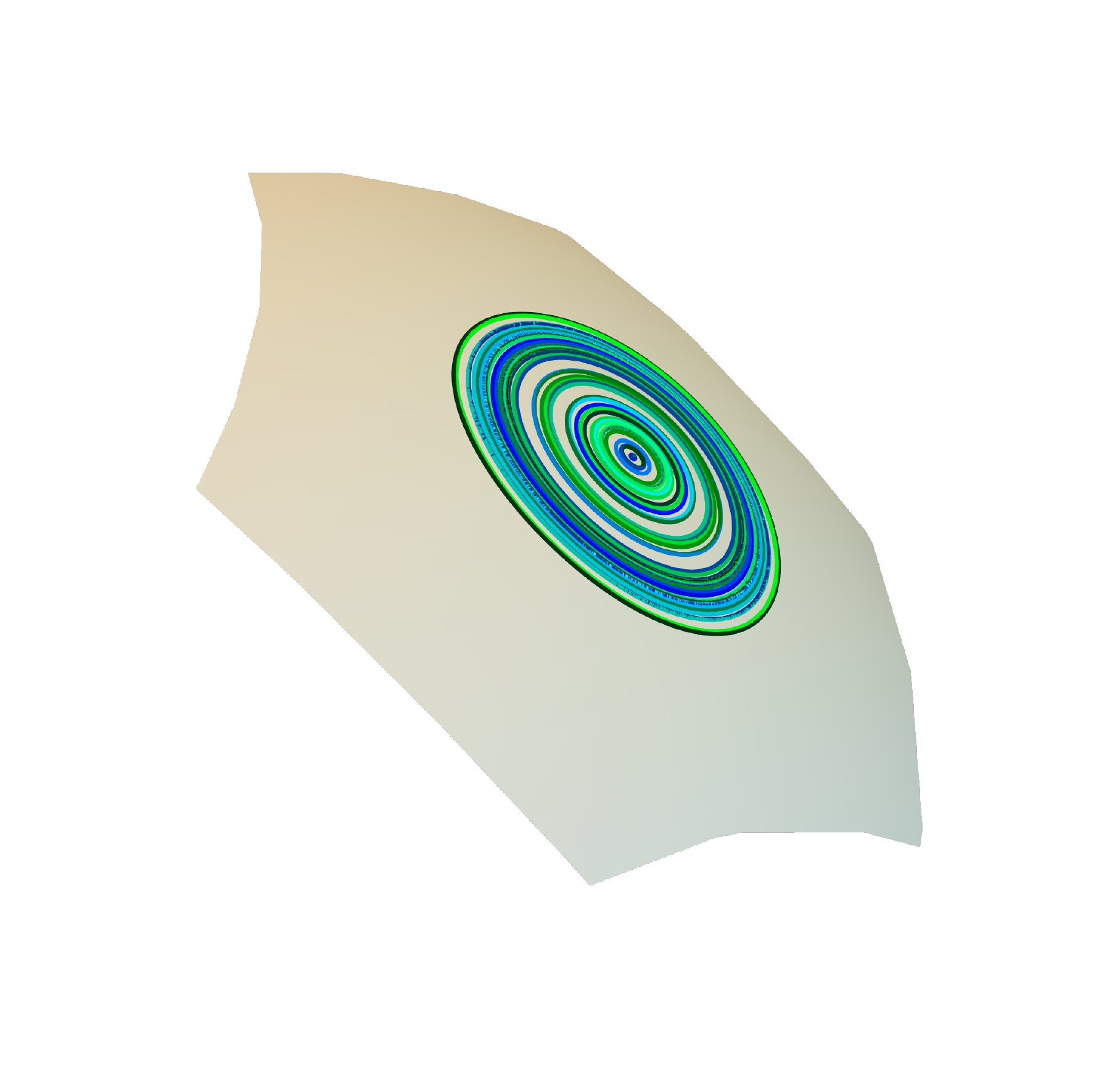}
\caption{A phase portrait of the dynamical system given by \cref{eqn:BracketReplicator} with $G = x_1x_2x_3$ and $H = x_1^2 + x_2^2 + x_3^2$ is shown. The payoff matrix is the ordinary RPS matrix. Evolution is shown on a portion of the surface of the 2-sphere.}
\label{fig:Sphere}
\end{figure}

Return now to the assumption that $G = x_1\cdots x_n$ and assume $\mathbf{A}$ is skew-symmetric. The Jacobi identity
\begin{equation}
\{F,\{H,L\}_{\mathbf{A},G}\}_{\mathbf{A},G} + \{H,\{L,F\}_{\mathbf{A},G}\}_{\mathbf{A},G} + \{L,\{F,H\}_{\mathbf{A},G}\}_{\mathbf{A},G} = 0.
\label{eqn:Jacobi}
\end{equation}
can be shown to hold for any skew-symmetric $\mathbf{A} \in \mathbb{R}^{n \times n}$ (a proof sketch is provided in the appendix). Thus for any skew-symmetric $\mathbf{A}$, the bracket
\begin{equation}
[F,H]_\mathbf{A} = \{F,H\}_{{x_1\cdots x_n},\mathbf{A}}
\end{equation}  
defines a Poisson algebra \cite{BV88} since the bracket satisfies (i) anti-commutativity, (ii) bilinearity, (iii) the Leibniz rule and (iv) the Jacobi identity. As noted, this bracket is distinct from the Poisson bracket and produces a corresponding geometry. We illustrate this with the simplest dynamic that does not occur on the simplex. Let $H = x_1^2 + x_2^2 + x_3^2$ and $\mathbf{A}$ be given by \cref{eqn:AMatrix0}. The result is a dynamical system that evolves on the sphere (see \cref{fig:Sphere}). There is no longer a convenient game-theoretic interpretation but the resulting dynamics clearly respect Liouville's theorem.
Studying the resulting geometry of this Poisson algebra is an area of future work.  

\section{Conclusions and Future Directions}
\textcolor{black}{Evolutionary games have been studied extensively by formulating Hamiltonians for the dynamics on the interior of the simplex. In early work, Akin and Losert \cite{AL84} show that zero-sum evolutionary games have a Hamiltonian structure. Because of that work's relationship to the work in this paper, we discuss it in more detail below. Hofbauer \cite{Hof96,HS98} studies bimatrix games as Hamiltonian systems. Hamiltonians are also used in the analysis of bimatrix games by Sato et al. \cite{SAF02,SAC05}, where chaotic behavior is analyzed, but no Poisson structures are considered. More recently, a Hamiltoninan approach is used in \cite{PT17} and \cite{AH18} to study extinction and thermodynamics in evolutionary games respectively.} 

\textcolor{black}{A more general (mathematical) study is conducted in \cite{AD14} in which a class of Poisson structures is introduced to study polymatrix games and applications to the ordinary replicator dynamics are provided. However this study does not consider those systems that satisfy Liouville's theorem and does not explicitly incorporate the results of Akin and Losert on zero-sum evolutionary games \cite{AL84}. Akin and Losert's analysis in \cite{AL84} is deep but non-intuitive. Though they construct a preserved volume form, the explicit metric is not stated and their proof requires an analysis of the foliations of the invariant manifolds of the dynamical system. In this paper,  we show that the replicator dynamic for zero-sum games emerges naturally from a parameterized bracket structure that is a hybrid between a Poisson and Nambu bracket. As a consequence we easily construct a metric in which phase space volume is preserved without resorting to symplectic geometry as in \cite{AL84}. This also allows us to characterize conditions when the ordinary Euclidean volume form is preserved, which is not considered in \cite{AL84}. Additionally, since it is known that dynamics arising from the Nambu bracket are degenerate Hamiltonian in higher-dimensions, it is likely that our analysis would provide a simpler mechanism for proving that zero-sum evolutionary games are Hamiltonian. We leave this as future work.} 

Moreover, because we have introduced a new Poisson algebra, our work expands both the dynamics that can be considered and suggests interesting potential work in relating Lie Algebras, Poisson Algebras and this new bracket. In particular, characterizing the infinitesimal generators of the corresponding Lie group would be of value. \textcolor{black}{We assume that these will be similar to the symplectomorphisms that characterize the Lie groups of the standard Poisson bracket.} Also, since the resulting bracket respects the Jacobi identity, it suggests a potential way to construct a quantized version of the replicator dynamic for zero-sum games, \textcolor{black}{thus extending work in quantum evolutionary games (see e.g., \cite{H06})}.

\section*{Acknowledgement}
Portions of CG's work were supported by the National Science Foundation Grant CMMI-1932991. C.G. would like to thank Andrew Belmonte for the discussion about this paper.

\newcommand{\pd}[2]{\frac{\partial#1}{\partial#2}}
\newcommand{\pdd}[3]{\frac{\partial^2#1}{\partial#2 \partial#3}}
\newcommand{\pdda}[2]{\frac{\partial^2#1}{\partial#2^2}}

\appendix
\section{Sketch of Proof of Jacobi Identity}\label{sec:A}
We first make use of the \cref{eqn:xixj} to simplify the definition of the bracket
\begin{equation}
[F,H]_\mathbf{A} = \sum_{i < j} A_{ij} x_i x_j\left(\frac{\partial F}{\partial x_i}\frac{\partial H}{\partial x_j} - \frac{\partial F}{\partial x_j}\frac{\partial H}{\partial x_i}\right).
\end{equation}
The similarity to the Poisson bracket is immediately clear and the proof that the bracket satisfies the Jacobi identity is identical to the proof that the Poisson bracket satisfies the Jacobi identity (see e.g., \cite{H13}) but with more terms because we must take into consideration the extra product with $x_i x_j$. The principle can be established in general by choosing an arbitrary term in e.g., $[L,[F,H]_\mathbf{A}]_\mathbf{A}$ and showing that it cancels with corresponding terms in $[F,[H,L]_\mathbf{A}]_\mathbf{A}$ and $[H,[L,F]_\mathbf{A}]_\mathbf{A}$, just as in the proof for the Poisson bracket. To see this, consider a single term in $[L,[F,H]_\mathbf{A}]_\mathbf{A}$
\begin{multline}
A_{12}x_1x_2\pd{L}{x_1}\left\{A_{12}\left[
x_1\left(\pd{F}{x_1}\pd{H}{x_2}-\pd{F}{x_2}\pd{H}{x_1}\right) + \right.\right.\\
\left.\left.
x_1x_2\left(\pdd{F}{x_1}{x_2}\pd{H}{x_2} + \pd{F}{x_1}\pdda{H}{x_2} - \pdda{F}{x_2}\pd{H}{x_1}-\pd{F}{x_2}\pdd{H}{x_1}{x_2}\right)
\right]
\cdots
\right\}
\label{eqn:BigTerm1}
\end{multline}
We now show two terms in $[F,[H,L]_\mathbf{A}]_\mathbf{A}$ that cancel some terms in \cref{eqn:BigTerm1}
\begin{multline*}
A_{12}x_1x_2\pd{F}{x_1}\left\{
A_{12}\left[
x_1\left(\pd{H}{x_1}\pd{L}{x_2} - \pd{H}{x_2}\pd{L}{x_1}\right) + \right. \right.\\
\left.\left.
x_1x_2\left(\pdd{H}{x_1}{x_2}\pd{L}{x_2} + \pd{H}{x_1}\pdda{L}{x_2} - \pdda{H}{x_2}\pd{L}{x_1}-\pd{H}{x_2}\pdd{L}{x_1}{x_2}\right)
\right] + \cdots\right.\\
\left.
A_{12}\left[
x_2\left(\pd{H}{x_1}\pd{L}{x_2} - \pd{H}{x_2}\pd{L}{x_1}\right) + \right. \right.\\
\left.\left.
x_1x_2\left(\pdda{H}{x_1}\pd{L}{x_2} + \pd{H}{x_1}\pdd{L}{x_1}{x_2} - \pdd{H}{x_1}{x_2}\pd{L}{x_1}-\pd{H}{x_2}\pdda{L}{x_1}\right)
\right]
\cdots
\right\}
\end{multline*}
and two terms in $[H,[L,F]_\mathbf{A}]_\mathbf{A}$ that cancel the remaining terms
\begin{multline*}
A_{12}x_1x_2\pd{H}{x_1}\left\{
A_{12}\left[
x_1\left(\pd{L}{x_1}\pd{F}{x_2} - \pd{L}{x_2}\pd{F}{x_1}\right) + \right. \right.\\
\left.\left.
x_1x_2\left(\pdd{L}{x_1}{x_2}\pd{F}{x_2} + \pd{L}{x_1}\pdda{F}{x_2} - \pdda{L}{x_2}\pd{F}{x_1}-\pd{L}{x_2}\pdd{F}{x_1}{x_2}\right)
\right] + \cdots\right.\\
\left.
A_{12}\left[
x_2\left(\pd{L}{x_1}\pd{F}{x_2} - \pd{L}{x_2}\pd{F}{x_1}\right) + \right. \right.\\
\left.\left.
x_1x_2\left(\pdda{L}{x_1}\pd{F}{x_2} + \pd{L}{x_1}\pdd{F}{x_1}{x_2} - \pdd{L}{x_1}{x_2}\pd{F}{x_1}-\pd{L}{x_2}\pdda{F}{x_1}\right)
\right]
\cdots
\right\}.
\end{multline*}
By symmetry, this argument applies to every term in the left-hand-side of the Jacobi identity, thus showing that equality with zero holds. We provide a Mathematica notebook in supplementary material to illustrate the cancellation of all terms in dimensions 3 and 4.

\bibliographystyle{unsrt}
\bibliography{NambuGamesBib}

\begin{thebibliography}{10}

\bibitem{AL84}
Ethan Akin and Viktor Losert.
\newblock Evolutionary dynamics of zero-sum games.
\newblock {\em J. Math. Bio.}, 20:231--258, 1984.

\bibitem{W97}
J.~W. Weibull.
\newblock {\em Evolutionary Game Theory}.
\newblock MIT Press, 1997.

\bibitem{HS98}
J.~Hofbauer and K.~Sigmund.
\newblock {\em {Evolutionary Games and Population Dynamics}}.
\newblock {Cambridge University Press}, 1998.

\bibitem{HS03}
J.~Hofbauer and K.~Sigmund.
\newblock {Evolutionary Game Dynamics}.
\newblock {\em Bulletin of the American Mathematical Society}, 40(4):479--519,
  2003.

\bibitem{T15}
Jun Tanimoto.
\newblock {\em Fundamentals of evolutionary game theory and its applications}.
\newblock Springer, 2015.

\bibitem{T19}
Jun Tanimoto.
\newblock {\em Evolutionary Games With Sociophysics}.
\newblock Springer, 2019.

\bibitem{Hof96}
J.~Hofbauer.
\newblock Evolutionary dynamics for bimatrix games: A {H}amiltonian system?
\newblock {\em J. Math. Bio.}, 34:675--688, 1996.

\bibitem{SAF02}
Yuzuru Sato, Eizo Akiyama, and J.~Doyne Farmer.
\newblock Chaos in learning a simple two-person game.
\newblock {\em Proceedings of the National Academy of Sciences},
  99(7):4748--4751, 2002.

\bibitem{SC03}
Yuzuru Sato and James~P Crutchfield.
\newblock Coupled replicator equations for the dynamics of learning in
  multiagent systems.
\newblock {\em Physical Review E}, 67(1):015206, 2003.

\bibitem{SAC05}
Yuzuru Sato, Eizo Akiyama, and James~P Crutchfield.
\newblock Stability and diversity in collective adaptation.
\newblock {\em Physica D: Nonlinear Phenomena}, 210(1-2):21--57, 2005.

\bibitem{RK20}
Vidya Raju and PS~Krishnaprasad.
\newblock Lie algebra structure of fitness and replicator control.
\newblock {\em arXiv preprint arXiv:2005.09792}, 2020.

\bibitem{GF17}
Christopher Griffin and James Fan.
\newblock Control problems with vanishing lie bracket arising from complete odd
  circulant evolutionary games.
\newblock {\em arXiv preprint arXiv:1710.09000}, 2017.

\bibitem{N73}
Yoichiro Nambu.
\newblock Generalized hamiltonian dynamics.
\newblock {\em Physical Review D}, 7(8):2405--2412, 1973.

\bibitem{EA83}
I.~Eshel and E.~Akin.
\newblock Coevolutionary instability of mixed nash solutions.
\newblock {\em J. Math. Bio.}, 18:123--133, 1983.

\bibitem{Ze80}
E.~C. Zeeman.
\newblock Population dynamics from game theory.
\newblock In {\em Global Theory of Dynamical Systems}, number 819 in Springer
  Lecture Notes in Mathematics. Springer, 1980.

\bibitem{BV88}
K.~H. Bhaskara and K.~Viswanath.
\newblock {\em Poisson algebras and Poisson manifolds}.
\newblock Longman, 1988.

\bibitem{PT17}
Hye~Jin Park and Arne Traulsen.
\newblock Extinction dynamics from metastable coexistences in an evolutionary
  game.
\newblock {\em Physical Review E}, 96(4):042412, 2017.

\bibitem{AH18}
Christoph Adami and Arend Hintze.
\newblock Thermodynamics of evolutionary games.
\newblock {\em Physical Review E}, 97(6):062136, 2018.

\bibitem{AD14}
Hassan~Najafi Alishah and Pedro Duarte.
\newblock Hamiltonian evolutionary games.
\newblock {\em arXiv preprint arXiv:1404.5900}, 2014.

\bibitem{H06}
Esteban~Guevara Hidalgo.
\newblock Quantum replicator dynamics.
\newblock {\em Physica A: statistical mechanics and its applications},
  369(2):393--407, 2006.

\bibitem{H13}
Brian~C Hall.
\newblock {\em Quantum theory for mathematicians}, volume 267.
\newblock Springer, 2013.

\end{thebibliography}
\end{document}